\documentstyle[aps,epsfig]{revtex}

\tighten
\begin{document}

\title{\bf Reheating in the Presence of Inhomogeneous Noise}

\author{V. Zanchin$^{1}$\footnote[1]{zanchin@super.ufsm.br}, 
A. Maia Jr.$^{2}$\footnote[2]{maia@ime.unicamp.br},  
W. Craig$^{3}$\footnote[3]{craigw@math.brown.edu} and
R. Brandenberger$^{4}$\footnote[4]{rhb@het.brown.edu}} 
  
\smallskip

\address{~\\$^1$Departamento de F\'{\i}sica, Universidade 
Federal de Santa Maria, \\97119-900, Santa Maria, RS, Brazil.}
 
\address{~\\$^2$Departamento de Matem\'atica Aplicada, \\
     Universidade Estadual de Campinas, 
     13083 - 970, Campinas, SP, Brazil.}

\address{~\\$^3$Mathematics Department and Lefschetz Center for 
Dynamical Systems,\\ Brown University, Providence, RI 02912, USA.}

\address{~\\$^4$Physics Department, Brown University, 
Providence, RI 02912, USA.}

\maketitle

\vskip 1.5cm
\begin{abstract}

\noindent 
Explosive particle production due to parametric resonance is a crucial 
feature of reheating in inflationary cosmology. Coherent oscillations 
of the inflaton field lead to a periodically varying mass in the evolution 
equation of matter and gravitational fluctuations and often induce a parametric resonance instability. In a previous paper (hep-ph/9709273) it was shown that homogeneous (i.e. space-independent) noise leads to an increase of the generalized Floquet exponent for all modes, at least if the noise is temporally uncorrelated. Here we extend the results to the physically more realistic case of spatially inhomogeneous noise. We demonstrate - modulo some mathematical fine points which are addressed in a companion paper - that the Floquet exponent is a non-decreasing function of the amplitude of the noise. We provide numerical evidence for an even stronger statement, namely that in the presence of inhomogeneous noise, the Floquet exponent of each mode is larger than the maximal Floquet exponent of the system in the absence of noise. 
\end{abstract}
 
\vfill

\setcounter{page}{0}
\thispagestyle{empty}

\vfill

\noindent BROWN-HET-1085 \hfill      December   1998.

\noindent hep-ph/yymmdd \hfill Typeset in REV\TeX

\vfill\eject

\baselineskip 24pt plus 2pt minus 2pt

\section{Introduction}

Explosive particle production at the end of the period of inflation is a crucial element of inflationary cosmology. It determines the effectiveness of energy transfer between the inflaton (the scalar field driving the exponential expansion of the Universe during inflation) and ordinary matter. As was first pointed out in \cite{TB} and discussed in detail in \cite{KLS94,STB,KLS97,GKLS97} and in many other papers (see e.g. \cite{others}), a coherently oscillating scalar field (such as the inflaton during the period of reheating) induces a parametric resonance instability in the mode equations of bosonic matter fields to which it couples. It has recently been realized that this resonance may also amplify gravitational waves \cite{Bassett}, scalar gravitational fluctuations \cite{FB98,PE98,BKM}, and - within the limits imposed by the exclusion principle - fermionic modes \cite{Baake,GK98}.

Resonance instabilities are, in general, quite sensitive to the presence or absence of noise. In a real physical system one expects some noise in the evolution of the scalar field driving the resonance. In the case of inflation, such noise could be due to quantum fluctuations (the same fluctuations which on larger length scales may develop into the seeds for galaxies and galaxy clusters) or to thermal excitations. In both cases, the amplitude of the noise is expected to be small. Nevertheless, it is important to study the effects of noise on the parametric resonance instability.

In a previous paper \cite{ZMCB97} we studied the effects of spatially homogeneous noise on the energy transfer between the inflaton $\phi$ and a second bosonic matter field $\chi$. We were able to prove that - provided the noise is temporally uncorrelated on time scales on the order of the period of oscillation of the inflaton - the presence of noise increases the rate of energy transfer between $\phi$ and $\chi$ (see also \cite{Bassett2} for similar results). More specifically, the exponential growth rate of a particular Fourier mode $\chi_k$ of $\chi$ is given by the generalized Floquet exponent $\mu_k$. Then, if the noise is given by a space-independent function $q(t)$, the following result holds \cite{ZMCB97}:
\begin{equation} \label{eq1}
   \mu_k (q) \, > \, \mu_k (0) \, \,\,\,\,\, \forall k \, ,
\end{equation}
if $q(t) \neq 0$. In particular, this implies that the stability bands (the bands of $k$ values for which $\mu(k) = 0$) disappear completely.

In this paper, we extend the above result to the physically more realistic case of spatially inhomogeneous noise. This extension involves introducing nonlinearities into the physical system. Inhomogeneous noise couples different Fourier modes of $\chi$. Hence, the mathematical analysis of the dynamics is much more complicated since the partial differential equation describing the evolution of $\chi$ can no longer be reduced via Fourier transformation to a system of decoupled ordinary differential equations. However, by introducing an ultraviolet cutoff and by considering the system in a finite volume, we can reduce the problem to a system of coupled ordinary differential equations. We are then able to show that the methods used in \cite{ZMCB97} extend to this case. Thus, for a given cutoff theory, the result (\ref{eq1}) can be shown to hold. When the cutoffs are removed, we obtain the slightly weaker result
\begin{equation} \label{eq2}
   \mu(q) \, \geq \, {\rm max}_k \mu_k (0) \, = \, \mu_{max} \, ,
\end{equation}
if $q(t,x) \neq 0$, where the Lyapunov exponent $\mu(q)$ is the maximal Lyapunov exponent of the inhomogeneous system. 

By Fourier transforming the configuration $\chi(x,t)$ at each time, it is possible to define a generalized Lyapunov exponent $\mu_k(q)$ for each Fourier mode. Mode mixing induced by the inhomogeneous noise has the effect of distributing the energy which in the absence of noise would be dumped only into the original instability bands to all of the modes of $\chi$. Hence, it is not unreasonable to expect that a much stronger result than (\ref{eq2}) holds, namely that $\mu_k(q)$ becomes larger than the maximal Floquet index $\mu_{max}$ of the system in the absence of noise. Indeed, we find numerical evidence which supports the following conjecture:
\begin{equation} \label{eq3}
   \mu_k (q) \, \geq \, \mu_{max} \, \,\,\,\,\, \forall k \, ,
\end{equation}
for a general inhomogeneous noise function $q(t,x)$.

The outline of this paper is as follows. In the following section we introduce the dynamical system to be studied and review the approach of \cite{ZMCB97} to proving (\ref{eq1}). In Section 3, the mathematically rigorous generalization of the proof to inhomogeneous noise is outlined, relegating some technical details to a companion paper \cite{CB98}. This yields a proof of (\ref{eq2}). In Section 4 we indicate how the main result can be seen via a perturbative approach, and in Section 5 we present numerical evidence for the much stronger statement of (\ref{eq3}). 
 
Our results apply for both systems with narrow and broad-band resonance. For simplicity we neglect the expansion of the Universe; however, we do not believe 
that including effects of expansion would change our main conclusions.   
 
\section{System with Homogeneous Noise}

The model we consider consists of two scalar fields $\phi$ and $\chi$ with an interaction Lagrangian of the form
\begin{equation} \label{intlag}
   {\cal L}_I \, \sim \, \epsilon \phi \chi^2 \, ,
\end{equation}
where $\epsilon \ll 1$ is a coefficient which parameterizes the strength of the interaction. 
In our application to inflationary cosmology, $\phi$ represents the inflaton field, the scalar field which drives inflation. When inflation ends, the inflaton begins to oscillate coherently about its ground state. Superimposed on this homogeneous oscillation is quantum or thermal noise $q(t, x)$, i.e.
\begin{equation}\label{phitx}
   \phi(t, x) \, = \, p(\omega t) + q(t, x) \, ,
\end{equation}
for some periodic function $p(\omega t)$.

We consider $\phi(t, x)$ to be a background field. Our goal is to study the effects of the resonant excitation of $\chi$ during the period when $\phi$ is oscillating coherently (this period is called the ``reheating" or more accurately the ``preheating" period \cite{KLS94}). The back-reaction of the excitation of $\chi$ on the evolution of $\phi$ can play an important role \cite{KLS97,GKLS97,others}. However, since this is not the subject of our study, we shall neglect this effect. The equation of motion for $\chi$ is
\begin{equation}
   {\ddot \chi} - \nabla^2 \chi + \left[m_{\chi}^2 + (p(\omega t) 
        + q(t, x))\right] \chi \,  = 0 \, ,
   \label{eq21}
\end{equation}
where $p(y)$ is a function with period $2 \pi$. Both the function $p(y)$ and the noise term in (\ref{eq21}) derive from the interaction Lagrangian (\ref{intlag}) (and are rescaled compared to (\ref{phitx})).

In the absence of noise, we can diagonalize the partial differential equation (\ref{eq21}) by Fourier transformation. The modes $\chi_k$ satisfy the following second order ordinary differential equation:
\begin{equation}
   {\ddot \chi_k} + \left[\omega_k^2 + p(\omega t) \right] \chi_k \, = 0 \, .
   \label{eq22}
\end{equation}
where $\omega_k^2 = m_{\chi}^2 + k^2$.
This is a Mathieu or Hill equation for each $k$, and as is well known,  there are (as a function of $k$) stability and instability bands. In the instability bands, $\chi_k$ grows exponentially at a characteristic rate given by the generalized Floquet index $\mu_k$. If $p / {\omega^2} \ll 1$, the instability bands are narrow, whereas in the case $p / {\omega^2} \gg 1$ we have broad-band resonance. Our analysis will apply to both cases.

In the case of homogeneous noise which was analyzed in \cite{ZMCB97}, the Fourier modes remain decoupled:
\begin{equation}
   {\ddot \chi_k} + \left[\omega_k^2 + p(\omega t) + q(t) \right] \chi_k \, = 0
   \, .
   \label{eq23}
\end{equation}
 It proves convenient to write the noise in terms of a dimensionless function  $n(t)$ with characteristic amplitude 1:
\begin{equation} \label{noise}
   q(t) = g \omega^2 n(t) \, , 
\end{equation}
where $g$ is a dimensionless coupling constant which is proportional to the coupling constant in (\ref{intlag}).

The first step in the analysis of \cite{ZMCB97} was to rewrite Equation (\ref{eq23}) for fixed $k$ (the index $k$ will be dropped in the rest of this section) as a first order matrix differential equation for the fundamental solution (or transfer) matrix $\Phi_q(\tau,0)$
\begin{equation} \label{matrixsol}
     \Phi_q(t, 0) \, = \, \left( \begin{array}{cc} 
                          \chi_1(t;q) & \chi_2(t;q) \\
                                   {\dot\chi_1(t;q)} & {\dot\chi_2(t;q)}
                         \end{array} \right) \, ,
\end{equation}
consisting of two independent solutions $\chi_1(t;q)$ and $\chi_2(t;q)$ 
of the second order equation (\ref{eq23}). The matrix equation reads
\begin{equation} \label{matrixeq}
   {\dot \Phi_q} \, = \, M(q(t), t) \Phi_q  \, ,
\end{equation}
with initial conditions $\Phi_q(0,0) = I$. Here, $M(q(t), t)$ is 
the matrix
\begin{equation} \label{matrix}
    M(q(t), t) \, = \, 
      \left( \begin{array}{cc}
      0 & 1  \\  
      - (\omega_k^2 + p(t) + q(t)) & 0 
       \end{array} \right) \, .
\end{equation}

For vanishing noise, i.e. $q(t) = 0$, the content of Floquet theory is
that the solution of (\ref{matrixeq}) can be written in the form
\begin{equation} \label{tm0matrix}
     \Phi_0(t, 0) \, = \, P_0(t) e^{Ct} \, ,
\end{equation}
where $P_0(t)$ is a periodic matrix function with period $T = {2 \pi}/\omega$, 
and $C$ is a constant matrix whose spectrum in a resonance region is 
$spec(C) = \{\pm \mu(0)\}$ with $\mu(0) > 0$.

The noise is supposed to describe quantum or thermal fluctuations. For the derivation which follows, it is sufficient to make certain statistical assumptions about $q(t)$. We assume that the noise is drawn from some sample space $\Omega$ (which for homogeneous noise can be taken to be $\Omega = C({\Re})$), and that the noise is ergodic, i.e. the time average of the noise is equal to the expectation value of the noise over the sample space. In this case, it can be shown that the generalized Floquet exponent of the solutions of (\ref{matrixeq}) is well defined by the limit
\begin{equation} \label{growth}
       \mu(q) \, = \, \lim_{N \rightarrow \infty} {1 \over {N T}} 
          \log \| \Pi_{j=1}^N \Phi(jT,(j-1)T) \| \, ,
\end{equation}
where $\| \cdot \|$ denotes some matrix norm (the dependence on the 
specific norm drops out in the large $N$ limit). Furthermore, the growth rate $\mu(q)$ is continuous in $q$ in an appropriate topology on $\Omega$ (see Appendix of \cite{ZMCB97} for details).

To obtain more quantitative information about $\mu(q)$ it is necessary to make further assumptions about the noise. We assume: 
\begin{description} 

\item{(i)} The noise $q(t)$ is uncorrelated in time on scales larger than $T$,
that is, $\{q(t; \kappa) \, : \, jT \leq t \leq (j + 1)T\}$ is independent of 
$\{q(t; \kappa) \, : \, lT \leq t \leq (l + 1)T\}$ for integers $l \neq j$, 
and is identically distributed., for all realizations $\kappa$ of the noise.

\item{(ii)} Restricting the noise $q(t; \kappa)$ to the time interval 
$0 \leq t < T$, the samples $\{q(t; \kappa): 0 \leq t < T\}$ within the 
support of the probability measure fill a neighborhood, in $C(0,T)$, of the 
origin.
\end{description}

Hypothesis (i) implies that the noise is ergodic, and therefore the 
generalized Floquet exponent is well defined. The main result which was proved in \cite{ZMCB97} is that  $\mu(q)$ is strictly larger than $\mu(0)$;
\begin{equation} \label{main}
     \mu(q) \, > \, \mu(0) \, ,
\end{equation}
which demonstrates that the presence of noise leads to a strict increase 
in the rate of particle production. The proof was based on an
application of Furstenberg's theorem, which concerns the Lyapunov exponent of products of independent identically distributed random matrices 
$\{\Psi_j : j=1, ... ,N\}$. 
 
Consider a probability distribution $dA$ on the matrices
$\Psi \in {\mathrm SL}(2n, {\mathrm R})$. Let $G_A$ be the smallest 
subgroup of ${\mathrm SL}(2n, {\mathrm R})$ containing the support of $dA$.

\medskip\noindent
{\bf Theorem 1: \ }{\it
(Furstenberg, \cite{pastur}) \ Suppose that $G_A$ is not compact, and that 
the action of $G_A$ on the set of lines in ${\mathrm R^{2n}}$ has no invariant 
measure. Then for almost all independent random sequences 
$\{ \Psi_j\}_{j=1}^\infty \subseteq {\mathrm SL}(2n, {\mathrm R})$ with
common distribution $dA$,
\[  
   \lim_{N\to\infty} {1 \over N} 
      \log (\| \Pi_{j=1}^N \Psi_j \|) = \lambda > 0 ~.
\]
Furthermore, for given $v_1, v_2 \in {\mathrm R^{2n}}$, then
\[  
    \lim_{N\to\infty} {1 \over N} 
      \log ( \langle v_1, \Pi_{j=1}^N \Psi_j v_2 \rangle ) = \lambda 
\] 
for almost every sequence $\{ \Psi_j\}_{j=1}^\infty$. 
}\medskip

In order to apply Furstenberg's theorem to obtain (\ref{main}), we start 
by factoring out from the transfer matrix $\Phi_q(t, 0)$ the contribution 
due to the evolution without noise;
\begin{equation} \label{factoriz}
     \Phi_q(t, 0) \, = \, \Phi_0(t, 0) \Psi_q(t, 0) \, = \, P_0(t) e^{Ct} 
     \Psi_q(t, 0) \, .
\end{equation}
The reduced transfer matrix $\Psi_q(t, 0)$ satisfies the following equation
\begin{equation} \label{Sdef}
     {\dot \Psi_q} \, = \, S(t; \kappa) \Psi_q \, = \, \Phi_0^{-1}(t)
     \left( \begin{array}{cc} 0 & 0 \\
             -q(t; \kappa) & 0
       \end{array} \right)\Phi_0(t) \Psi_q \, ,
\end{equation}
which can be written as a matrix integral equation
\begin{equation} \label{redint}
     \Psi_q(t, 0) \, = \, 
         I + \int_0^t d\tau \ S(\tau; \kappa) \Psi_q(\tau, 0) \, .
\end{equation}
By properties (i) of decorrelation of the noise, the quantities
$\Psi_q(jT, (j-1)T)$ are independent and identically distributed 
for different integers $j$, and we can  apply the Furstenberg theorem 
to the following decomposition of $\Psi_q(NT, 0)$;
\begin{equation}
     \Psi_q(NT, 0) \, = \, \Pi_{j=1}^N \Psi_q(jT, (j-1)T) \, .
\end{equation}
We may choose for instance the vector $v_1$ in Theorem 1 to be an 
eigenvector of $\Phi_0(T,0)^t = (P_0(T) e^{CT})^t$, the transpose of the 
transfer matrix of the system without noise, with eigenvalue $e^{\mu(0)T}$. 
Then the second statement of Theorem 1 becomes 
\begin{eqnarray}
  {1 \over {NT}} \log |\langle v_1, \Phi_q(NT, 0) v_2 \rangle | 
  & = & {1 \over {NT}} \log 
        |\langle v_1, P_0(NT) e^{CNT} \Psi_q(NT, 0) v_2 \rangle | 
  \nonumber \\
  & = & {1 \over {NT}} \log ( e^{\mu(0)NT} 
        |\langle v_1, \Psi_q(NT, 0) v_2 \rangle |) \\
  & = & \mu(0) \, + \, {1 \over {NT}} \log 
        |\langle v_1, \Pi_{j=1}^N \Psi_q(jT, (j-1)T) v_2 \rangle | \, .
  \nonumber
\end{eqnarray}
Taking the limit $N \rightarrow \infty$ and applying the first and second statements of Theorem 1 we obtain
\begin{equation}
   \mu(q) \, = \, \mu(0) + \lambda \, > \, \mu(0)
\end{equation}
which proves the main result (\ref{main}). Note in particular that 
(\ref{main}) implies that the stability bands of the system without noise disappear when spatially homogeneous noise is added.

\section{Exact Results}

In this section we will generalize the results of \cite{ZMCB97} to the case of inhomogeneous noise, modulo some technical points which will be addressed in a separate paper \cite{CB98}. To keep the notation simple, we will take space to be one-dimensional. However, the analysis works in any spatial dimension. The starting point is the second order differential equation (\ref{eq21}) which we can rewrite as a first order matrix operator equation
\begin{equation} \label{matrixeq2}
   {\dot \Phi_q} \, = \, M(q(t,\cdot), t) \Phi_q  \, 
\end{equation}
with initial condition $\Phi_q(0) = I$ for the fundamental solution matrix $\Phi_q(t)$ with kernel $\Phi_q(t;x,y)$ given by
\begin{equation} \label{matrixsol2}
     \Phi_q(t;x,y) \, = \, \left( \begin{array}{cc} 
                          \chi_1(t;x,y) & \chi_2(t;x,y) \\
                                   {\dot\chi_1(t;x,y)} & {\dot\chi_2(t;x,y)}
                         \end{array} \right) \, ,
\end{equation}
consisting of two independent kernels $\chi_1(t;x,y)$ and $\chi_2(t;x,y)$ 
of the linear operator equation (\ref{eq21}). Here, $M(q(t,x), t)$ is the matrix operator
\begin{equation} \label{matrix2}
    M(q(t,x), t) \, = \, 
      \left( \begin{array}{cc}
      0 & 1  \\  
      - (-\nabla^2 + m_{\chi}^2 + p(t) + q(t,x)) & 0 
       \end{array} \right) \, .
\end{equation} 
The operators are taken to act on the Hilbert space $H = H^1({\mathrm R^3}) \times L^2({\mathrm R^3})$, where $H^1$ stands for the Sobolev space of $L^2$ functions with $L^2$ gradients. The first factor
refers to the coordinate function, the second to the velocities. Physically, our choice of function space corresponds to considering finite energy configurations.

As in the previous section, we introduce rotating coordinates in order to factor out the time evolution in the absence of noise:
\begin{equation}
     \Phi_q(t) \, = \, \Psi_q(t) \Phi_0(t) \, ,
\end{equation}
where $\Phi_0(t)$ is the fundamental solution matrix operator in the absence of noise, and the product implies operator composition. The initial conditions are $\Psi_q(0) = I$. In Fourier space this is a generalization of the matrix given in the previous section (\ref{tm0matrix}). The kernel of the matrix operator $\Psi_q(t)$ satisfies the reduced differential equation 
\begin{equation} \label{Sdef2}
     {\dot \Psi_q(t;x,y)} \, = \, \int S(t;x,z_1;q) \Psi_q(t;z_1,y) dz_1 \, = \, \int \int \Phi_0^{-1}(t;x,z_2)
     \left( \begin{array}{cc} 0 & 0 \\
             -q(t,z_2) & 0
       \end{array} \right)\Phi_0(t;z_2,z_1) \Psi_q(t;z_1,y) dz_1 dz_2 \, .
\end{equation}
The corresponding operator equation can also be written as an integral equation
\begin{equation} \label{inteq2}
     \Psi_q(t) \, = \, I + \int_0^t dt'S(t';q) \Psi_q(t') \, .
\end{equation}

We will focus on the reduced transfer matrix operator
\begin{equation}
     \Psi_q(NT) \, = \, \Pi_{j=1}^N \Psi_q(jT,(j-1)T)
\end{equation}
which describes the evolution from $t = 0$ to $t = NT$, where $T$ is the period of the oscillating source $p(t)$, factored into matrix operators $\Psi_q(jT,(j-1)T)$ which describe the evolution in each individual period. By differentiating (\ref{inteq2}) with respect to the noise function $q$ we obtain the action of the transfer matrix operator in the Lie algebra of infinitesimal displacements $r(t,x)$
\begin{equation} \label{infinit}
   \delta \Psi_q(T) \cdot r = 
   \int_0^T \ dt \ \  \Phi_0^{-1}(t) 
         \left( \begin{array}{cc} 
            0  & 0 \\  
            -r(t,\cdot) & 0   
         \end{array} \right) \Phi_0(t)  ~.
\end{equation}
This construction will be used in the derivation of our main results.

The first relevant mathematical result is that the growth rate (or generalized Floquet or Lyapunov exponent) of the solutions of (\ref{matrixeq2}) is well defined by the limit
\begin{equation}
     \mu(q) \, = \, \lim_{N \rightarrow \infty} {1 \over {N T}} 
          \log \| \Pi_{j=1}^N \Phi(jT,(j-1)T) \| \, ,
\end{equation}
where $\| \cdot \|$ denotes the operator matrix norm 
\begin{equation}
\| \Phi \| \, = \, sup_{v \in H^1 \times L^2} \| \Phi(v) \|_H \, ,
\end{equation}
where the subscript $H$ indicates that the function norm in the Hilbert space $H$ is used. Furthermore, the growth rate $\mu(q)$ is continuous in $q$ in an appropriate topology on the noise sample space. Both of these results were already demonstrated in the Appendix of \cite{ZMCB97} to which the reader is referred for details.

The Lyapunov exponent of the wave equation without noise, i.e. of the wave operator $\Phi_0(t)$ is given by
\begin{equation}
     \mu_0 \, = \, \lim_{t \rightarrow \infty} {1 \over t} 
          \log \| \Phi_0(t) \| \, = \, max_{k \in R^3} \mu_k(0) \, ,
\end{equation}
where $\mu_k(0)$ is the Lyapunov exponent of the k'th mode of the system without noise.

We assume that the noise function $q(t,x)$ is given by a stochastic process which is temporally uncorrelated over times $t$ larger than the period $T$. In this case, the quantities $\Psi_q(jT,(j-1)T)$ are identically distributed uncorrelated random matrix operators. The main theorem of this paper is the following:

\medskip\noindent
{\bf Theorem 2: \ }  \  When a spatially inhomogeneous noise $q(t,x;\omega)$ is introduced into the problem, then the Lyapunov exponent can only increase, i.e.
\[
     \mu(q) \, \geq \, \mu_0
\]
for almost all realization $\omega$ of the noise.
\medskip

The main idea of the proof is to reduce the infinite dimensional operator problem to a finite dimensional matrix problem to which Furstenberg's Theorem in the form stated in Section 2 can be applied. First, we introduce an infrared cutoff by putting the system in a box. This corresponds to a discretization of the Fourier modes. The separation of k values is denoted by $\Delta k$. Next, we impose an ultraviolet cutoff in momentum space by eliminating all modes with frequency larger than $k_{max}$. This reduces the problem to a finite dimensional case to which Furstenberg's Theorem applies. We will prove that the hypotheses of the theorem are indeed satisfied.

In order to be able to take the limit when the ultraviolet cutoff goes to infinity and the infrared cutoff to zero, it is important to prove that the  transfer matrix $\Psi_q$ in the continuum system is a compact operator perturbation of the transfer matrix operator without noise. This implies that the perturbation can be well approximated by the corresponding transfer matrix operators of the finite dimensional approximate problem obtained after imposing the cutoffs. Since a product of compact operators is compact, it is sufficient to prove that both $\Phi_0(T,x) - \Phi_0(T,x)_{| p=0}$ and $\Psi_q(T,x) - I$ are compact independently (the first term describes the difference in the transfer matrix when the periodic part of the perturbation, namely $p(t)$, is turned on, the second term expresses the effect of non-vanishing $q(x,t)$).
  
\medskip\noindent
{\bf Theorem 3: \ }  \  The operators $\Phi_0(T,x) - \Phi_0(T,x)_{| p=0}$ and $\Psi_q(T,x) - I$ are compact.
\medskip

The compactness of $\Phi_0(T,x) - \Phi_0(T,x)_{| p=0}$ follows via the known asymptotics of the eigenfunctions of Sturm-Liouville operators. The compactness of $\Psi_q(T,x) - I$ can be shown explicitly \cite{CB98}.

Given the infrared and ultraviolet cutoffs introduced above, it is natural to work in Fourier space. In this basis, the partial differential equation (\ref{eq21}) for $\chi(t,x)$ becomes a system of $2n$ coupled ordinary differential equations for the Fourier components $\chi_k(t)$ and their momenta ${\dot \chi_k(t)}$, where $n$ is the number of independent Fourier modes. In other words, the transfer matrix equation (\ref{matrixeq2}) becomes a $2n$ dimensional vector equation with a $2n \times 2n$ transfer matrix $\Phi_q(t)$.
If we order the basis vectors such that
\begin{equation}
    \chi \, = \, \left( \begin{array}{cc} 
            \chi_{k_1}  &   \\  
            {\dot \chi_{k_1}} & \\
            ... & \\
            \chi_{k_n} & \\
            {\dot \chi_{k_n}} &    
         \end{array} \right) \, ,
\end{equation}
where $k_1, ..., k_n$ label the finite set of momenta we are considering, then $\Phi_0(t,x)$, the transfer matrix in the absence of noise, is block diagonal
\begin{equation}
\Phi_0(t,x) \, = \, \left( \begin{array}{ccc} 
            \Phi_{0,k_1}  & 0 & 0  \\  
            0 & ... & \\
            0 & 0 & \Phi_{0,k_n}     
           \end{array} \right) \, .
\end{equation}
Here, the matrices $\Phi_{0,k_i}$ are the $2 \times 2$ matrices for the $k_i$'th mode in the absence of noise (see Section 2). The matrix $S$ of (\ref{Sdef2}), on the other hand, mixes the blocks. If we define a new matrix $Q$ by
\begin{equation}
S(t,x;q) \, = \, \Phi_0(t,x)^{-1} Q(t,x;q) \Phi_0(t,x) 
\end{equation}
and denote the Fourier transform of $q(t,x)$ by ${\tilde q}(t,k)$, then
\begin{equation}
Q(t,x;q) \, = \, \left( \begin{array}{cccc} 
            Q_{11}  & Q_{12} & .. & Q_{1n}  \\  
            Q_{21}  & Q_{22} & .. & Q_{2n}  \\
            ..      & ..     & .. & ..      \\
            Q_{n1}  & Q_{n2} & .. & Q_{nn}      
            \end{array} \right) \, ,
\end{equation}
where $Q_{ij}$ is the $2 \times 2$ matrix
\begin{equation}
Q_{ij}(t,k_i - k_j) \, = \, \left( \begin{array}{cc} 
            0  & 0   \\  
            {\tilde q}(t,k_i - k_j)  & 0        
            \end{array} \right) \, .
\end{equation}

We are now ready to sketch the proof of the main theorem. It is sufficient to show that the hypotheses of the Furstenberg Theorem (see Section 2) are satisfied. Let us denote
\begin{equation}
\Psi_j \, = \, \Psi_q(jT,(j-1)T) \,\,\,\,\, \in SL(2n, R) \, .
\end{equation}
and consider $G_A$, the smallest subgroup of $SL(2n, R)$ containing the support of $dA$, the set of transfer matrices which can be obtained from realizations of the noise drawn from the noise sample space. We must show that

\begin{description}
\item{(i)} $G_A$ is not compact.
\item{(ii)} $G_A$ acting on the set of lines in $R^{2n}$ has no invariant measure, i.e. there is no subspace of $R^{2n}$ left invariant under the action of all of the transfer matrices corresponding to noise drawn from the sample space.
\end{description}

To show that $G_A$ is not compact, it is sufficient to consider the subset $H_A \subset R^{2n}$ corresponding to homogeneous noise and to show that $H_A$ is not compact. Thus, we take
\begin{equation}
{\tilde q}(t,k_i - k_j) \, = \, q(t) \delta(k_i - k_j) \, .
\end{equation}
Thus, $H_A$ is block diagonal and does not mix the basis vectors corresponding to different values of $k$. Since the set $\{ q(t) \}$ is not compact, it follows that within each block, $H_A$ is non-compact.

Turning to the second hypothesis of Furstenberg's Theorem, it follows from our previous study of homogeneous noise (see Appendix of \cite{ZMCB97}) that no subspace of $R^{2n}$ corresponding to a fixed value of $k$ is invariant under the action of $H_A$ and hence of $G_A$.
To show that there cannot be an invariant subspace which is different from a subspace of an $R^2$ associated with an individual value of $k$, we argue by contradiction. It is convenient to do the analysis at the level of the Lie algebra associated with the group of transfer matrices, namely the set of infinitesimal displacements given by (\ref{infinit}), and then to use the implicit function theorem to recover the corresponding result at the group level.
 
Assuming that there exists an invariant subspace $X \subset R^{2n}$, we construct a special noise function $q(t,x)$ such that the associated transfer matrix takes some vector $v \in X$ into a vector $w \not\in X$. For example, let us assume that the subspace $X$ contains a vector $v \in R_{\alpha}^2$ associated with the $\alpha$'th Fourier mode, but that it is orthogonal to a vector $w \in R_{\beta}^2$ associated with the $\beta$'th Fourier mode. Consider now a noise function which is defined by
\begin{equation}
{\tilde q}(t,k_i - k_j) \, = \, q(t) \delta(k_i - k_{\alpha}) \delta(k_j - k_{\beta})\, .
\end{equation}
Under this noise function, the vector $v$ is mapped onto a vector $v'$ which is not orthogonal to $w$. The generalization of this argument to arbitrary vectors $v$ and $w$ will be given in \cite{CB98}.

\section{Born Approximation}
  
In this section we follow our previous work 
\cite{ZMCB97} closely. 
The inflaton field is taken to be a real scalar field and is 
denoted by $\phi$. In the period immediately after inflation, 
$\phi(t)$ is assumed to be oscillating coherently about 
(one of) its ground state(s). As in the previous section, we shall, however, include a  small aperiodic,
inhomogeneous perturbation (cf. eq. (\ref{phitx})).

We shall (following \cite{TB,KLS94,STB}) take $\phi$ to be 
coupled to a second scalar field $\chi$ which represents 
matter via an interaction Lagrangian which is quadratic 
in $\chi$, for example of the form (\ref{intlag}).
To simplify the discussion, we neglect nonlinearities in the 
equation of motion for $\chi$. Such nonlinearities may  
 be important and lead to an early termination of 
parametric resonance. This topic has recently been 
discussed extensively, see e.g. \cite{KLS97,others}, 
but is not the topic of our paper. Instead, we are interested 
whether the presence of noise such as included in 
Eq. (\ref{phitx}) will affect the onset of resonance.

For simplicity, we will neglect the expansion of the Universe. 
In \cite{TB,KLS94,STB} it has been shown that it is possible to include the expansion without any problems and that it does not 
affect the results concerning the parametric resonance 
instability. Since the equation of motion for $\chi(x,t)$ 
is linear, we can  write down the evolution equation for 
the Fourier modes of $\chi$,
denoted by $\chi_{k}(t)$
or, equivalently, by $\chi({\bf k},t)$. 
The equation of motion for the $\chi$
field is given by (\ref{eq21}), where $q(t,{\bf x})$ represents 
the noise which we consider as a perturbation of the driving function $p(t)$.
 Taking the Fourier Transform of  equation  (\ref{eq21}) and using the
convolution theorem we obtain the equation of evolution for
the $k$-mode
\begin{equation}
\ddot {\chi}_{k} + \left[\omega_{k}^2+ p(\omega t)\right] \chi_{k} 
= -  \int d^3 {\bf k}^{'} q({\bf k' - k},t) \chi({\bf k'},t)
\label{eqn2}\end{equation}
where
$\omega_{k}^2 = m_{\chi}^2 + {\bf k}^2 $ ,
and $\chi_{k}\equiv \chi({\bf k},t)$ is the Fourier transform of
 $\chi({\bf x},t)$. 

Note that all modes on the right hand  side of Eq. (\ref{eqn2}) will 
influence the time evolution of the particular
$k$-mode we are considering. This is a very important aspect
of our analysis for the case of inhomogeneous noise.

A straightforward calculation shows that (\ref{eqn2}) is 
equivalent to an integral equation (resembling  Volterra's equation),
namely

\begin{equation}
\chi_{k}(t) = \chi_k^{(h)} + \chi_k^{(1)} = \chi_{k}^{(h)} - \int^{t}_{t_{i}} dt' G(t,t')
\int d^{3}{\bf k'} q({\bf k' - k},t) \chi({\bf k'},t)
\label{eqn4}
\end{equation}
where $t_i$ is the time at the beginning of reheating and $\chi^{(h)}_{k}$ is the general solution
 of the ``homogeneous equation"
\begin{equation}
\ddot {\chi} + \left[{\bf k}^2+ m_{\chi}^2 +  p(\omega t)\right] \chi = 0 .
\label{eqn5}
\end{equation}

\noindent For the particular case when $q({\bf k'-k},t)= q(t)
\delta({\bf k'-k})$ we obtain the equation discussed in
\cite{ZMCB97}.

Since $p(\omega t)$ is a periodic function, $\chi^{(h)}_{k}$
can be written in the Floquet form
\begin{equation}
\chi^{(h)}_{k} =  e^{\mu_{k} t} p_{1}(t) + 
 e^{-\mu_{k} t} p_{2} (t) \label{eqn6}
\end{equation}
where $p_{1}(t)$ and $p_{2}(t)$ are periodic functions (which include the arbitrary
constants determined by initial conditions) and 
$\mu_{k}$ is the Floquet exponent.

The solutions $\chi_{1}(t) = e^{\mu_{k} t} p_{1}(t)$ and
$\chi_{2} = e^{-\mu_{k} t} p_{2}(t) $
are linearly independent and we can choose them in such a way  that 
their Wronskian
$W = \chi_{1}(t) \dot{\chi_{2}}(t) - \dot{\chi_{1}}(t) 
\chi_{2}(t)$ is time independent (see e.g. \cite{Rabenstein}). Then, the kernel $G(t,t')$ 
in Eq.(\ref{eqn4})  is given by
\begin{equation}
G(t,t') = {\chi_{1}(t') {\chi_{2}}(t) -{\chi_{1}}(t)\chi_{2}(t') \over W }
\label{eqn7}
\end{equation}

Our method, in this section, consists of finding an approximate 
solution of the integral equation (\ref{eqn4}), by reducing it
as closely as possible to the equation in the case of homogeneous noise
previously  obtained in \cite{ZMCB97}.
In the first approximation we replace $\chi({\bf k'},t)$ in the
momentum-space integral in (\ref{eqn4}) by the homogeneous solution
$\chi^{(h)}({\bf k'},t)$. The integral then reduces to
\begin{equation}
\int d^{3} {\bf k'} q({\bf k' - k},t) p_{1}({\bf k'},t)
e^{\mu_{k} t} + \int d^{3} {\bf k'} q({\bf k' - k},t)
 p_{2}({\bf k'},t)
e^{-\mu_{k} t} 
\label{eqn8}
\end{equation}

If we restrict to the case of narrow resonance, the
main contribution to the above integral comes from the first
resonance band, since higher bands are of the order 
$o(\epsilon^{2})$ or higher. In this case the integrals in the above
expression can be replaced by integrals over a finite domain, namely, the
first resonance band which, in momentum space, is given by the shell
 $k_{1} \le |{\bf k'}| \le k_{2}$. For our analysis 
we do not need to know the values of $k_{1}$ and $k_{2}$. The
expression (\ref{eqn8}) above reduces to
\begin{equation}
\int^{|{\bf k}| = k_{2}}_{|{\bf k}| = k_{1}} d^{3} {\bf k'} 
q({\bf k' - k},t) p_{1}({\bf k'},t)
e^{\mu_{k} t}  + \int^{|{\bf k}| = k_{2}}_{|{\bf k}| = 
k_{1}} d^{3} {\bf k'} q({\bf k' - k},t)
 p_{2}({\bf k'},t)
e^{-\mu_{k} t }
\label{eqn9}
\end{equation}

If we assume that the periodic function $p_{2}$ has an 
amplitude of the same order as (or less than) the function $p_{1}$,
then we can neglect the integral
containing $p_{2}$, since its integrand is exponentially
decreasing. Thus we are left with only the first integral in
(\ref{eqn9}). Now, for this integral, we apply the following
Mean Value Theorem \cite{theo}:

{\bf Theorem} (Mean Value in 3 Dim): 
Let $f,g:V \subset {\cal R}^{3}
\longrightarrow {\cal R}$ be continuous functions, where $V$ is
a compact domain in ${\cal R}^{3}$. Then there exists a
$\overline{x} \in V$ such that
\begin{equation}
\int_{V} f({\bf x})g({\bf x}) d^{3}{\bf x} = f(\overline{\bf x}) 
\int_{V} g({\bf x}) d^{3}{\bf x} \, . \label{eqn10}
\end{equation}

Using this theorem, (\ref{eqn9}) reduces to
\begin{equation}
e^{\mu_{\overline{k}} t}
\int^{|{\bf k}| = k_{2}}_{|{\bf k}| = k_{1}} d^{3} {\bf k'} 
q({\bf k' - k},t) p_{1}({\bf k'},t) \equiv 
e^{\mu_{\overline{k}} t} A({\bf k},t) \, ,
\label{eqn11}
\end{equation}
where $\overline{{\bf k}}$ is inside  the resonance band.
This is our final expression which estimates the contribution 
of all modes in the evolution equation of the mode ${\bf k}$.
So, using the above result, our earlier evolution equation
(\ref{eqn2}) reads
\begin{equation}
\ddot {\chi_{k}} + \left[\omega_{k}^2 + (p(\omega t) \right] \chi_{k} = 
- A({\bf k},t) e^{\mu_{\overline{k}} t}  \label{eqn12}
\end{equation}

A closer analysis of the function $A({\bf k},t)$, defined 
in (\ref{eqn11}), shows that it is bounded. This is because
$p_{1}$ is periodic and the noise is by assumption a random function with a small
amplitude which is identically distributed over time periods of $T$.
Thus, we can use the approximation
\begin{equation} \label{approx}
|A({\bf k},t)| \le  M_{1} \sigma_{k} \omega^3 \epsilon \, ,
\end{equation}
where $M_{1}$ and $\sigma_{k}$  are, respectively, upper bounds of the functions
$p_{1}({\bf k'},t)$ and $q({\bf k'-k},t)$ as both variables 
${\bf k'}$ and $t$ are varied. The last two factors in (\ref{approx}) represent an estimate of the 
volume of the first (and dominant) instability band. Also, let $M_{2}$ be the upper bound
on $p_{2}({\bf k'},t)$.
 \begin{figure}
\begin{center}
   \mbox{\epsfig{figure=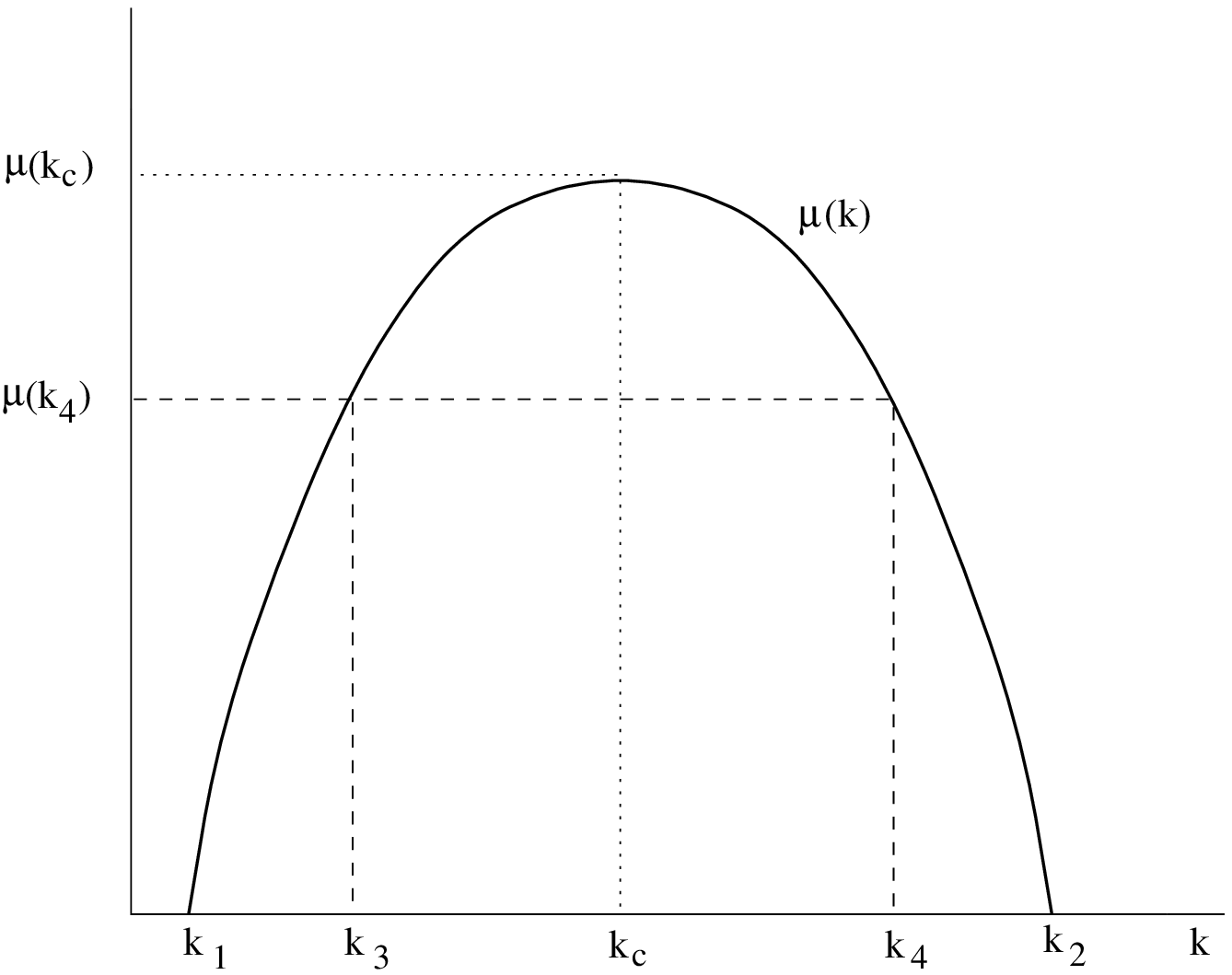,height=2.5in}}
\end{center}
\caption{ The Floquet exponent $\mu$  is plotted  (vertical axis)
against $k/\omega$ (horizontal axis), for a general case of the Hill equation}
 \end{figure}

In the resonance band $[k_{1},k_{2}]$, we must consider
the Floquet exponent as a function of $k$. The exact formula 
for $\mu_{k}$ can be complicated, but it can be approximated very well 
by a circle of arc centered at the
midpoint of the interval $[k_{1},k_{2}]$.
Let $k_{c}$ denote the central mode with the
maximum exponent. Note that $\mu(k_{1})=\mu(k_{2})=0$ (see Fig. 1). Then, if
$\overline{k} \neq k_{c}$, we have
$\mu_{\overline{k}} \le \mu_{k}$, for all $k$  satisfying
$|k-k_{c}| \le |\overline{k} - k_{c}|$. We will now consider the subinterval
$[k_{3},k_{4}] \subset [k_{1},k_{2}]$ ``centered"
at $k_{c}$ for which this is the case, and we shall show that for values of $k$ in this
subinterval the noise does not destroy the parametric resonance.

From
Eqs. (\ref{eqn4}) and (\ref{eqn11}) it follows that
\begin{equation}
\chi^{(1)}_{k}(t) = - \int^{t}_{t_{i}} dt' G(t,t')
A({\bf k},t') e^{\mu_{\overline{k}}t'}
\label{eqn14}
\end{equation}
where the Green's kernel $G(t,t')$ is given by (\ref{eqn7}). Using (\ref{approx}) it can
easily be shown that 
\begin{equation}
|\chi^{(1)}_{k}| \le \sigma_{k} f(k,\Delta t) 
e^{\mu_{k} \Delta t}
\label{eqn15}
\end{equation}
where $\Delta t = t-t_{i}$ and the function $f(k,\Delta t)$ 
is given by
\begin{equation}
f({\bf k},\Delta t) =
M_{1}^{2} M_{2} W^{-1} e^{\mu_{k} t_{i}}
(1- e^{- \mu_{k} \Delta t}) \epsilon \omega^3 {1 \over {\mu_k}}
\label{eqn16}
\end{equation}
Comparing (\ref{eqn15}) with the growth rate of the homogeneous
solution $\chi^{(h)}_{k}$, that is $e^{ \mu_{k} \Delta t}$, we
find
\begin{equation}
{|\chi^{(1)}_{k}| \over e^{ \mu_{k} \Delta t}} \le
  \sigma_{k} f(k,\Delta t)
\label{eqn17}
\end{equation}
for all ${\bf k}$ in the subinterval $k_{3} \le |{\bf k}| \le
k_{4}$ inside the resonance band (a shell in the 3-dim momentum
space) $[k_{1},k_{2}]$.

The above inequality shows that small amplitude noise (which can be
controlled by the parameter $\sigma_{k}$) will not change
the exponential rate of growth in the parametric resonance
regime.

This method applies only under very restrictive hypotheses 
such as the assumption of narrow resonance band dynamics, applicability of the
First Born Approximation method, and yields only a result in a
subinterval of the resonance band which may  be much narrower
than the band itself (in fact, for $\overline{k} = k_{c}$, only for
the central mode $k_{c}$ we could guarantee that noise does not
eliminate the exponential growth). However, based on the numerical results to be
presented in the following section 
we gain  some confidence that the methods could be generalized to less
restrictive hypotheses (perhaps including the case of broad resonance). 
This will be the subject of further studies. A very important
aspect in our analysis is the fact that the noise considered
here is quite general. This is good since there are many possible sources
of noise during and before the end of the inflationary period. 

\section{Numerical Results} 
   
In order to estimate the effects of inhomogeneous noise on the resonant
field through numerical approximations we take the ``discrete" Fourier
Transform of equation (\ref{eq21}) to get
\begin{equation}
\ddot {\chi}_{\bf k}(t) + \left[\omega_{k}^2+ 
p(\omega t)\right] \chi_{\bf k}(t) = - \sum_{\bf k'} 
q_{{\bf k'} - {\bf k}}(t) \chi_{\bf k'}(t)\, ,\label{eqnr1} 
\end{equation}
where $\chi({\bf x},t) = \sum_{\bf k}\chi_{\bf k}(t) \exp[i{\bf k.x}]$, etc.

   Let us  analyze the time evolution of a given mode ${\bf k}$.  We can 
write (\ref{eqnr1}) in the following form
\begin{equation}
\ddot {\chi}_{\bf k}(t) + \left[\omega_{k}^2+  p(\omega t)+ 
 q_o(t)\right] \chi_{\bf k}(t) =  F_{\bf k}\, ,\label{eqnr2}
\end{equation}
where $F_{\bf k} \equiv -  
\sum_{{\bf k'}\neq {\bf k}} q_{{\bf k'} - {\bf k}}(t) \chi_{\bf k'}(t)$
does not contain $\chi_{\bf k}$ and then may be thought of as an external force.

   The general solution to (\ref{eqnr2}) has the form
\begin{equation} \label{eqnr3}
\chi_{\bf k} = \chi_{\bf k}^h +\chi_{\bf k}^p\, ,
\end{equation}
where $\chi_{\bf k}^h$ is the solution to the homogeneous equation 
\begin{equation}
\ddot {\chi}_{\bf k}(t) + \left[\omega_{k}^2+ 
p(\omega t)+ q_o(t)\right] \chi_{\bf k}(t) =  0\, ,
\label{eqnr4}\end{equation}
and $\chi_{\bf k}^p$ is the particular solution of (\ref{eqnr2})
\begin{equation}\label{eqnr5}
\chi_{\bf k}^p(t)= \int_{t_o}^t G(t,t')F_{\bf k}(t') dt'\, .
\end{equation}
$G(t,t')$ is the Green function associated to (\ref{eqnr4})
\begin{equation}
G(t,t') = {\chi_{1}(t'){\chi_{2}}(t) - {\chi_{1}}(t)\chi_{2}(t') \over W } \, ,
\label{eqnr6}
\end{equation}
where $\chi_1$ and $\chi_2$ are the two independent solutions of (\ref{eqnr4}) and
$W$ is the Wronskian, which in the present case is a constant.

Equation (\ref{eqnr4}) is the same as the evolution equation for the 
${\bf k}$-mode of the boson field coupled to the inflaton during the 
pre-heating phase of the Universe and in the presence of homogeneous noise.  
This problem was studied
at length in \cite{ZMCB97}. According to the results obtained in that analysis,
the presence of ergodic noise strictly increases the rate of 
particle production.
In other words, the generalized Floquet exponent in the presence of noise 
$\mu_k(q_o)$ is greater than $\mu_k(0)$, the Floquet exponent for 
$q_o(t)=0$.  We can write $\mu_k(q_o) = \mu_k(0) + \lambda_k(q_o)$ with
$\lambda_k(q_o) >0$.
We then have two distinct situations to analyze depending upon whether 
the ${\bf k}$ mode is in a resonance band or not.  Modes in a stability
band have $\mu_k(0) =0$, while the modes in a resonance band have
$\mu_k(0) >0$.

To simplify the discussion we neglect the contribution of 
$q_o(t)$, i.e. set $\lambda_k(q_o) = 0$  in Eq. (\ref{eqnr4}). 
This assumption will not change 
the qualitative results derived below. Moreover, for the situations 
we are interested in, where the noise is in general small
when compared to the amplitude of the inflaton, and for modes close to
the center of the resonance band, $\lambda_k(q_o)$ is small
compared to $\mu_k(0)$ and can be neglected.
 
  Let us first assume that ${\bf k}$ is in a resonance band. Then according 
to the theory of the Hill equation, the solution to (\ref{eqnr4}) with 
$q_o(t)=0$ is 
\begin{equation}  \label{eqnr7}
 \chi_{\bf k}^h(t) =e^{\mu_k(0) t}p_1(t) + e^{-\mu_k(0) t} p_2(t)\,
\end{equation}
where $p_1$  and $p_2$ are bounded (periodic) functions of time.  
The solution to (\ref{eqnr2}) for the modes within any resonance band
 and for late times can be written as
\begin{equation}\label{eqnr8}
\chi_{\bf k}(t) \simeq e^{\mu_k(0) t}p_1(t) + \chi_{\bf k}^p(t)\, .
\end{equation}
{}From the above result we see that $\chi_{\bf k}(t)$ has an exponentially 
growing component independent of $\chi_{\bf k}^p(t)$. Therefore, the 
exponential growth of a mode in a resonance band can only be destroyed by 
inhomogeneous noise if there is an  exact cancellation between
$\chi_{\bf k}^h(t)$ and  $\chi_{\bf k}^p(t)$. This is only possible if $\chi_{\bf k}^p(t)$
is given by (at least for late times) $- e^{\mu_k(0) t}p_1(t)$, which is very unlikely since 
the initial conditions for all the 
modes are involved (see below).  Notice that $\chi_{\bf k}^p(t)$ can in fact
grow exponentially in time and reinforce the increase of $\chi_k(t)$, since all modes are 
coupled through the term on the right hand side of (\ref{eqnr1}).  The 
result is a stimulated resonance effect on the particular components, even 
for modes in a stability band (see below).

  We now study the time evolution of non-resonant modes.  Our goal is to 
estimate the rate of growth of such modes. 
For the sake of simplicity, we take $q_o(t) \simeq 0$ as above. 
Therefore,
the homogeneous solution $\chi_{\bf k}^h(t)$ is given by two independent
periodic functions $\chi_1(t)$ and $\chi_2(t)$. As we are 
interested in the exponentially
growing components, we are left with only the particular solution 
$\chi_{\bf k}^p(t)$ to compute.  In the following we proceed to find an 
upper bound for this ``particular" solution. The results hold for all 
modes in a stability band.
 In order to estimate this solution, which
 is given by (\ref{eqnr5}), we first need an approximation for $F_{\bf k}(t)$.
Let us then explicitly determine the contribution from the homogeneous 
and particular solutions of each mode. Using (\ref{eqnr3}) and the 
definition of $F_{\bf k}(t)$, we can write
\begin{equation}\label{eqnr9}
F_{\bf k}(t)=- \sum_{{\bf k}'\neq {\bf k}}
q_{{\bf k}'-{\bf k}}(t) \left(e^{\mu_{k'}(0) t}P_{{\bf k}'}(t)
+ e^{-\mu_{k'}(0) t}Q_{{\bf k}'}(t)+\chi_{{\bf k}'}^p(t)\right)\, ,
\end{equation}
where $P_{\bf k}(t)$ and $Q_{\bf k}(t)$ are periodic functions of $t$. Let us recall that $\mu_{k}(0)$, the Floquet exponent, is (according to our 
definition) zero for modes in a stability band and positive for resonant modes.
   To begin with, we neglect bounded and decaying components in the 
above equation and keep just exponentially growing terms.  
 Thus, using (\ref{eqnr6}) and recalling that in the present case
the solution $\chi_1$ and $\chi_2$ that enter the Green's function
$G(t,t')$ are periodic functions of $t$, we can write
\begin{eqnarray}\label{eqnr10}
\chi_{\bf k}^p(t)&=&-\sum_{{\bf k}' \in B} \int_{t_o}^t G(t,t')
q_{{\bf k}'-{\bf k}}(t') \, e^{\mu_{k'}(0) t'}P_{\bf k}'(t)\,
-\sum_{{\bf k}'\neq {\bf k}}\int_{t_o}^t G(t,t')q_{{\bf k}'-{\bf k}}(t')\chi_{{\bf k}'}^p(t') \nonumber \\
& &+ \mbox{terms not growing exponentially in time}\, ,
\end{eqnarray}
where $B$ stands for the sub-sets of the ${\bf k}$-space ${\mathbf R^3}$, such that ${\bf k}$ is in a resonance band. Continuing our approximations, we consider the fastest growing term
on the right had side of  (\ref{eqnr10}). Let $k_c$ denote such a mode so 
we can cast (\ref{eqnr10}) in the form
\begin{eqnarray}\label{eqnr11}
\chi_{\bf k}^p(t)&=&- \int_{t_o}^t G(t,t')
q_{{\bf k}_c-{\bf k}}(t') \,e^{\mu_{k_c}(0) t'}P_{{\bf k}_c}(t)\,
- \sum_{{\bf k}'\neq {\bf k}}\int_{t_o}^t G(t,t')q_{{\bf k}'-{\bf k}}(t')\chi_{{\bf k}'}^p(t') + R\, ,
\end{eqnarray}
where $R$ stands for terms that grow slower than $e^{\mu_{k_c}(0) t}$.
The first term in the above equation guarantees  an exponential growth of
$\chi_{\bf k}^p(t)$ with the same rate as the fastest growing resonant 
mode. It is also worth noting that this leading term does not depend on 
the initial condition of the mode itself, but just on the initial conditions
of the mode ${\bf k}_c$. This result holds for every mode in a stability band
and is confirmed by numerical calculations as depicted in Figure 2(a).
 
Note that in  equation (\ref{eqnr11})
${\bf k}_c \neq {\bf k}$, since we assumed that ${\bf k}$ is in a stability 
band while ${\bf k}_c$ is the central mode of the broadest resonance band
of the specific model considered.  In a narrow resonance band regime 
${\bf k}_c$ corresponds to the central mode of the first resonance band. 
In a broad resonance regime, however, ${\bf k}_c$ may not be the central mode,
but it is still the mode which has the fastest exponential growth. For instance,
in the model studied in \cite{KLS97}, ${\bf k}_c =0$ is the mode with 
the greatest Floquet exponent.

 To verify the previous qualitative results we performed a numerical analysis
of the problem. 
Following the notation of Ref. \cite{ZMCB97}, we choose $p(t)=
\lambda\cos(2 t)$, where $\lambda$ is a constant, and the noise given in the form 
$q_{\bf k}(t)=g\,\lambda\,n(t)\,m(k)$. Here, $g$ is a positive 
constant, $n(t)$ is a random function of time with characteristic
time $T = \Gamma_t^{-1}$ and amplitude equal to unity, and 
$m(k)$, also 
a random function with characteristic rate $\Gamma_k$ and amplitude 1,
depends only upon the magnitude of ${\bf k}$. 
We use the dimensionless time $t$ ( $=\omega t$) so that  
the evolution equation for $\chi_k$ reads
\begin{equation}
\ddot{\chi}_k +\left[E_k^2 + \lambda \cos(2t)\right]\chi_k =
\lambda\, g\, n(t) \sum_{k'}\chi_{k'} m_{k'-k} \, ,
\label{eqnr12} \end{equation}
where  $\lambda$ and $E_k^2$ are normalized  with respect to the frequency
$\omega$ of the inflaton field. Let us stress that the amplitude of the 
noise was chosen to be very small, in such a way that all the corresponding
homogeneous samples of the noise would not change the solution 
significantly. 
This choice assures that the effects we are about to show
are consequence solely of the inhomogeneity of the noise.

 Figure 2 shows a particular case of broad resonance where $E_k^2$ is of the
form $E_k^2 = \lambda+ k^2$ ($k$ is also in units of $\omega$), with $E_o^2 =\lambda=30$.  In this particular case, $m_{k}$ is chosen to be a 
smooth function of $k$ which helps us to show that the growth of modes
outside a resonance band does not depend upon their initial conditions 
(see below).
\begin{figure}
\begin{center}
   \mbox{\epsfig{figure=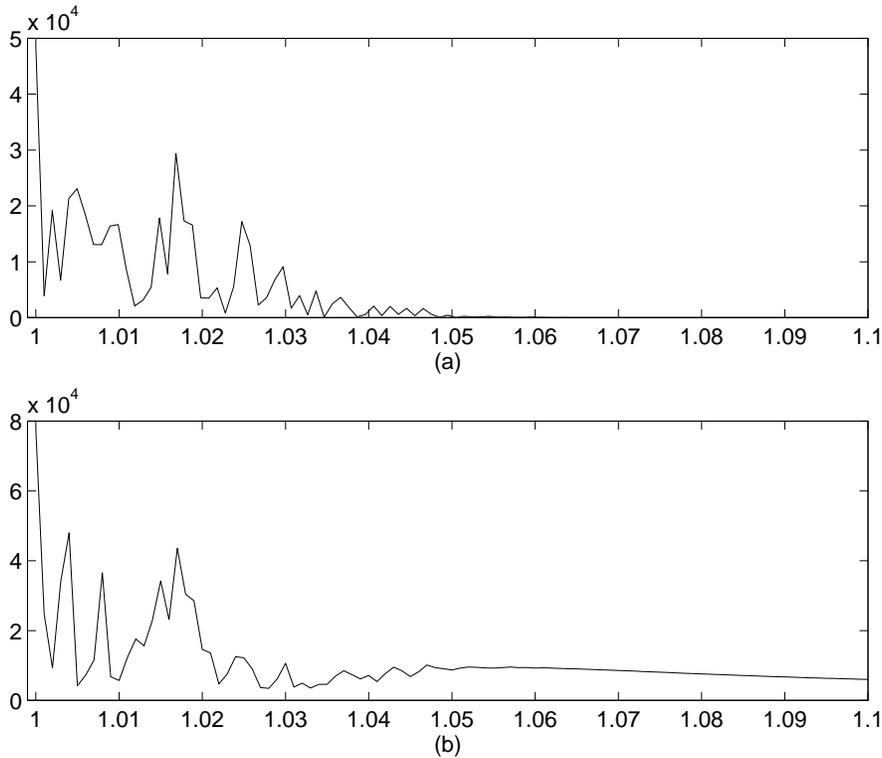,height=4.0in}}
\end{center}
\caption{The maximum amplitude of $\chi_k(t)$ in $t\in [0,40]$ is plotted  
(vertical axis) against  $E_k^2/\lambda$ (horizontal axis), $E_o^2 =\lambda=30$.\\
(a) No noise is present, $g=0$.\\
(b) The noise is characterized by $\Gamma_t = 10$ and $g = 0.3$ 
(small amplitude and large noise frequency).}
\end{figure}

  This figure shows clearly the stimulated resonance of the modes in a
stability band. Fig. 2(a) shows a particular case of broad resonance in 
the absence of noise. The random peaks arise because we have chosen 
random  initial conditions for both $\chi_k(t)$ and $\dot{\chi}_k(t)$ 
at $t=0$.  In Fig 2(a) the ratio between the amplitudes of the modes 
in the stability band (for $E_k^2/\lambda > 1.05$) and the central 
mode is $O(10^{-4})$, while in Fig 2(b), which shows the results 
when inhomogeneous noise is present, that ratio is $O(10^{-1})$.
 Observe that in Fig 2(b) all modes outside the resonance band have almost 
the same amplitude. The absence of peaks in that region demonstrates the 
independence of the amplitudes on the initial conditions of such modes. 
Those amplitudes depend strongly on the initial conditions and Floquet 
exponent of the central mode $\chi_{k_c}(t)$.  Only $100$ modes were taken 
into account in this example.
\begin{figure}
\begin{center}
   \mbox{\epsfig{figure=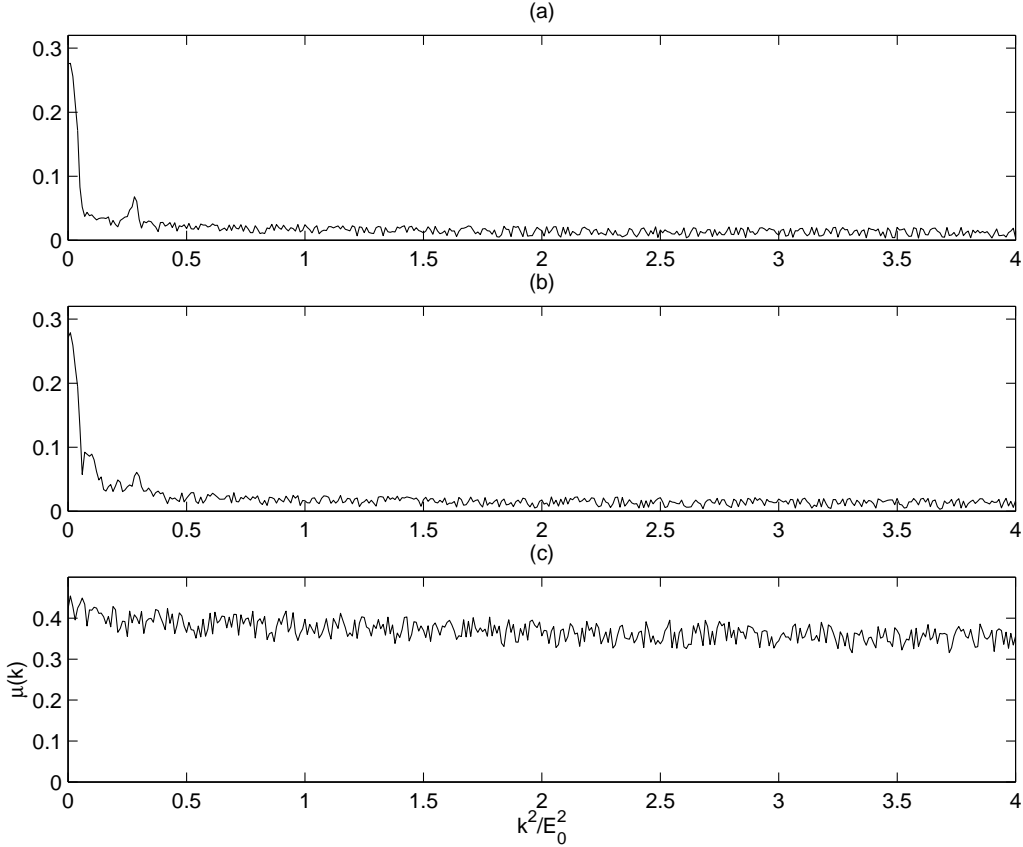,height=4.5in}}
\end{center}
\caption{The Floquet index $\mu(k)$  is plotted  
(vertical axis) against  $k^2/E_0^2$ (horizontal axis), $E_o^2 =\lambda=30$.\\
(a) No noise is present, $g=0$.\\
(b) The noise is characterized by $\Gamma_t = 10$ and $g = 0.1$ 
The noise is homogeneous ($m_{k'-k}=\delta(k'-k)$).\\
(c) Same parameters as in (b). The noise is inhomogeneous.}
\end{figure}

   Figures 3(a)--3(c) show even more explicitly the effects of inhomogeneities 
in the noise. In this case $k$ is normalized in such a way that it assumes
$400$ different values between $k_0 =0$ end $k_{max} = 2 E_0$. In 3(a) no
noise is present and we can identify the first resonance band. Figure 3(b)
shows the result when we introduced homogeneous noise. The noise is random 
in both $t$ and $k$ and the amplitude is small, $g = 0.1$.  Very
small effects due to the noise are observed. Only modes within the resonance
 band grow exponentially, exactly as in the case of Fig. 3(a).
On the other hand, in Fig. 3(c) we observe a very distinct behavior of
the modes outside the resonance gap. All the modes have explosive growth, 
and we cannot distinguish the stability region from the resonance band
anymore. The inhomogeneity of the noise has large effects even in the case
studied here,
where the amplitude of the noise is small compared to the amplitude of the potential $p(t)$. 

\begin{figure}
\begin{center}
   \mbox{\epsfig{figure=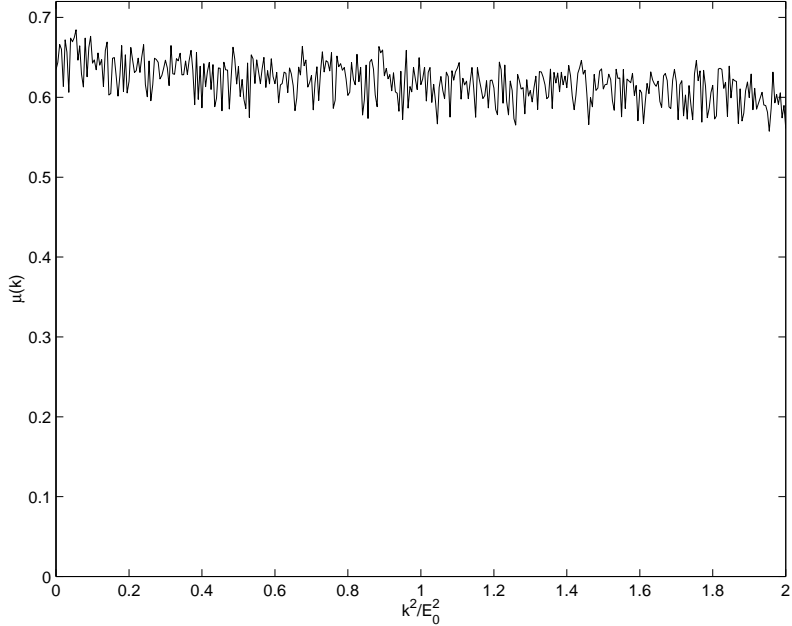,height=3.5in}}
\end{center}
\caption{The same as figure 3(c), but here with $\Delta k$ twice smaller
than in 3(c).}
\end{figure}

 Another interesting point is related to the abovementioned spacing
$\Delta k$ between two successive modes.  We can easily see how the
Floquet exponent depends on $\Delta k$ by  comparing
Fig. 3(c) with Fig. 4. In the last case we used the same parameters as in
3(c), only the normalization of $k$ is different: $400$ modes are 
placed between $k_{min}=0$ and $k_{max} = \sqrt{2}E_0$.  $m_{k'-k}$ was 
chosen to be a symmetric (otherwise random) matrix in $k$ and $k'$ in both the cases
of Fig. 3(c) and Fig. 4.  The Floquet exponent increases when the
number of Fourier modes is increased. This can be understood since it
corresponds to decreasing $\Delta k$ between two 
neighboring modes. As a result, we are adding the contribution of a larger 
number of very close modes which are inside the resonance band, all
of them contributing with exponentially growing amplitudes, on the right 
hand side of Eq. (\ref{eqnr12}). 

\section{Conclusions}

In this paper we have shown (modulo some technical points which will be demonstrated
in a companion paper) that spatially inhomogeneous noise in the oscillating field which induces the
parametric resonance instability does not decrease the Floquet exponent of the instability. This
extends the results of \cite{ZMCB97} where it was shown that homogeneous noise leads to an increase
in the Floquet exponent for each Fourier mode. Crucial to our proof was the truncation of the
dynamical system to a finite dimensional one by means of infrared and ultraviolet cutoffs, use
of the Furstenberg Theorem concerning the Lyapunov exponent of products of identically distributed
random matrices, and the use of compactness arguments to carry over the main results as the
cutoffs are removed.

We have also provided numerical evidence (backed up with approximate analytical calculations) which
indicate that a much stronger result is true, namely that inhomogeneous noise will spread the
exponential growth of the most resonant mode of the system without noise to all of the Fourier
modes. An interesting challenge is to provide a mathematically rigorous proof of this stronger
result.

\section*{Acknowledgments}

This work is partially supported by CNPq and FAPESP (Brazilian Research 
Agencies), the Divison of Mathematical Sciences of the U.S. National Science Foundation and by the U.S. Department of Energy under contract DE-FG0291ER40688,
Task A .


\begin{thebibliography}{30}
\bibitem{theo} This is a generalization of the Mean Value Theorem to three dimensions.
\end{thebibliography}
\end{document}